\documentclass[10pt, conference, compsocconf]{IEEEtran}

\def\IEEEbibitemsep{0pt plus .3pt}
\makeatletter
\def\ps@IEEEtitlepagestyle{
  \def\@oddfoot{\mycopyrightnotice}
  \def\@evenfoot{}
}

\def\mycopyrightnotice{
  {\footnotesize
  \begin{minipage}{\textwidth}
  \centering
  Copyright~\copyright~2020 IEEE.  Personal use of this material is permitted.  Permission from IEEE must be obtained for all other uses, in any current or future media, including reprinting/republishing this material for advertising or promotional purposes, creating new collective works, for resale or redistribution to servers or lists, or reuse of any copyrighted component of this work in other works.
  \end{minipage}
  }
}

\ifCLASSINFOpdf
\else
\fi

\usepackage{cite}
\usepackage{amsmath,amssymb,amsfonts}

\usepackage{graphicx}
\graphicspath{ {images/} }
\usepackage[caption=false]{subfig}
\usepackage{url}

\usepackage[outdir=./]{epstopdf}
\usepackage{textcomp}
\usepackage{xcolor}

\let\labelindent\relax

\usepackage{enumitem}

\usepackage{algorithm}
\usepackage{algorithmicx}
\usepackage[noend]{algpseudocode}

\usepackage[utf8]{inputenc}
\usepackage[english]{babel}
\usepackage{lipsum}

\usepackage{amsthm}
 
\theoremstyle{definition}

\begin{document}
%
\title{ Signature-based Selection of IaaS Cloud Services \vspace{-7mm} }


\author{\IEEEauthorblockN{Sheik Mohammad Mostakim Fattah\IEEEauthorrefmark{1}, Athman Bouguettaya\IEEEauthorrefmark{1}, and Sajib Mistry\IEEEauthorrefmark{2} }
\IEEEauthorblockA{School of Computer Science, University of Sydney, Australia\IEEEauthorrefmark{1}\\
                  School of Electrical Engineering, Computing and Mathematical Sciences, Curtin University, Australia\IEEEauthorrefmark{2}\\
Email: \{sheik.fattah, athman.bouguettaya\}@sydney.edu.au\IEEEauthorrefmark{1}, sajib.mistry@curtin.edu.au\IEEEauthorrefmark{2}} 
}

\maketitle

\begin{abstract}
We propose a novel approach to select IaaS cloud services for a long-term period where the service providers offer \textit{limited} QoS information. The proposed approach leverages free short-term \textit{trials} to obtain the previously undisclosed QoS information. A new significance-based trial scheme is proposed using frequency distribution analysis to test a consumer's long-term workloads in a short trial. We introduce a novel IaaS \textit{signature} technique to uniquely identify the variability of a provider's QoS performance. A Signature-based QoS Performance Discovery (SPD) algorithm is proposed which leverages the combination of free trials and IaaS signatures. A set of exhaustive experiments with real-world datasets is conducted to evaluate the proposed approach.




\end{abstract}


%
\IEEEpeerreviewmaketitle

\section{Introduction}

Infrastructure-as-a-Service (IaaS) is a key service delivery model that offers virtualized computational resources in the cloud market \cite{iosup2014iaas}. Virtual Machines (VMs), Virtual Storage (VS), and Virtual Private Networks (VPNs) are some common services provided through IaaS models. IaaS cloud offers an effective alternative to manage an organization's in-house IT infrastructure in the cloud. Amazon, Google, and Microsoft are examples of leading IaaS providers. IaaS providers often promote \textit{long-term} services (i.e., 1 - 3 years) by offering significant discounts. For example, Amazon offers up to 75\% discounts for the reserved EC2 instances compared to on-demand instances\footnote{\url{https://aws.amazon.com/ec2/pricing/reserved-instances/}}.


Large organizations such as airline companies, banks, and research institutes tend to utilize IaaS services on a \textit{long-term basis} for economic reasons. Selecting the right IaaS service is an important \textit{business decision} for \textit{long-term consumers} \cite{ye2016long}. Consumers usually determine \textit{long-term service requirements} based on their expected revenue, market expansion, history, and budget \cite{mistry2018metaheuristic}. 

IaaS models utilize the service paradigm as a mechanism to deliver services \cite{fattah2019long}. An IaaS service consists of two parts: \textit{functional} and \textit{non-functional}. Functional attributes are set based on the purpose of the service such as computing, data storing, and networking. Non-functional attributes are the Quality of Service (QoS) attributes such as availability, response time, and throughput. QoS attributes help a consumer to select the \textit{best performing} services from a large number of functionally similar services \cite{yu2010computing}. 

IaaS providers typically do not provide adequate information about their service performance to make an \textit{informed} selection \cite{fattah2018cp}. Selecting the best performing IaaS service is challenging due to the \textit{incomplete IaaS advertisements} and \textit{limited performance history}. We address these challenges using a signature-based IaaS selection approach. The proposed approach predicts service performance by utilizing the patterns of service performance behavior as represented by a provider's performance signature. We identify the following key challenges in the long-term IaaS selection:

   \noindent \textbf{a) Incomplete advertisements}: IaaS providers reveal \textit{limited} and \textit{short-term} QoS information in their advertisements \cite{wang2018testing}. IaaS advertisements typically contain a limited number of QoS attributes. For instance, disk read/write throughput, memory bandwidth, and availability are unavailable in most advertisements \cite{iosup2011performance}. The advertised performance information may not be representative for a long period. For instance, a consumer may want to know the performance in December where the advertised performance is recorded in June. Additionally, the advertised information may not be \textit{helpful} to understand service performance due to the lack of detailed information. For example, EC2 instances have different types of virtual CPUs (vCPUs). Each vCPU can be a thread of an Intel Xeon core, an AMD EPYC core, or AWS Graviton processor according to AWS advertisements\footnote{\url{https://aws.amazon.com/ec2/instance-types/}}. Estimating the vCPU's actual performance is difficult from such limited information \cite{feitelson2002workload}. Providers often advertise an average or maximum performance of their services. For instance, the network performance of some EC2 instances has up to a 10-gigabit data transfer rate. Existing studies show that providers often fail to offer the \textit{promised} QoS performance in the long-term period \cite{ye2016long}.
   
    \noindent \textbf{b) Limited performance history}: IaaS Providers usually do not share detailed service performance history publicly due to \textit{market competition} and \textit{business secrecy} \cite{fattah2019long}. There exist third party data collectors such as CloudHarmony, and CloudSpectator that provide summarized results or insights on the performance of cloud services. These results are usually not fit for further analysis due to the reduced dimensions in QoS attributes and time \cite{li2010cloudcmp}. For instance, CloudHarmony mainly monitors network availability and does not provide any insight on response time and throughput. Moreover, collectors often use proprietary benchmarks but reveal limited information about the benchmarking process.

Existing approaches to select IaaS services with incomplete information leverage \textit{free short-term trials} advertised by IaaS providers \cite{wang2018testing,ye2016long}. For instance, Microsoft offers \$200 credits to explore any Azure service for 30 days. A consumer may discover the short-term performance of different QoS attributes based on its workloads in a free trial. An equivalence-partitioning based trial strategy is proposed to discover a provider's long-term service performance \cite{fattah2019long}. The proposed approach focuses on the temporal restriction of short-term trials during the trial workload generation. The following aspects of the long-term selection using free trials have not been addressed:


    \noindent \textbf{a) Workload characteristics}: A consumer may execute a wide variety of workloads over a long period. A service may exhibit \textit{inconsistent} performance behavior for different types of workloads \cite{wang2018testing}. For example, a service may exhibit higher throughput for CPU-intensive workloads than network-intensive workloads. Free trials are typically offered for a short period (e.g., 7 to 30 days). Effective utilization of free trials is a prerequisite to make an informed selection. It may be challenging to test all kinds of workloads in a short period. Hence, \textit{trial workloads should be selected carefully to maximize the utilization of free trials.}
    
    \noindent \textbf{b) Performance variability}: Commercial IaaS providers typically operate in a \textit{multi-tenant} environment and over-commit resources. A provider's service performance may fluctuate based on several factors such as active co-tenants, QoS management policy, and location \cite{wang2018testing}. It is therefore \textit{inadequate to rely only on} the trial experience for a long-term commitment. Most existing long-term selection approaches assume that the long-term service performance of a provider is known \cite{liu2015qos,fattah2019long}. In practice, a consumer does not know the provider's long-term service performance.

We propose a \textit{novel approach} that leverages free trials to select IaaS services for a long-term period according to a consumer's QoS requirements. The approach introduces two new concepts: a) \textit{IaaS signature} which captures the long-term IaaS performance variability, and b) \textit{workload significance} which addresses a consumer's future workload characteristics. The contributions are summarized as follows:


\begin{itemize}[itemsep=0ex, leftmargin=*]
    \item A significance-based trial scheme to discover the unknown QoS performance for a consumer's long-term workloads.
    \item An IaaS signature technique to uniquely identify a provider's QoS performance variability. 
    \item A signature-based IaaS selection approach that utilizes the trial experience and IaaS signatures to discover long-term IaaS performance.
\end{itemize}

\section{Related Work}
\label{sec:related}

IaaS cloud selection is a topical research challenge in cloud computing \cite{jayasinghe2012expertus}. Several IaaS selection approaches are proposed to find the optimal IaaS providers based on their QoS performance. A cloud comparison approach called CloudCom is proposed to help consumers select a cloud provider that fits their needs \cite{li2010cloudcmp}. CloudCom addresses three key services, i.e., elastic computing, persistent storage, and networking services. The performance of each service is measured based on the \textit{most relevant QoS attributes} that may affect consumers' applications directly. The IaaS service selection problem is often modeled as a multi-criteria decision-making problem \cite{liu2015qos}. We categorize the exiting IaaS service selection approaches into the following groups:

\noindent\textbf{a) Short-term IaaS selection}: A common approach to select IaaS services is to perform short-term trials using a representative application or micro-benchmarks \cite{wang2018testing}. Existing studies suggest that traditional benchmarks for computer systems are not suitable for cloud performance discovery \cite{iosup2014iaas}. A generator approach is proposed to automate performance testing in IaaS cloud \cite{jayasinghe2012expertus}. The proposed work aims at reducing human errors for large scale distributed experiments. Several studies suggest that the IaaS performance fluctuates considerably based on the application workloads \cite{feitelson2002workload}. These studies focus on short-term selection and do not consider the long-term performance change.

\noindent\textbf{b) Long-term IaaS selection}: The long-term IaaS selection approach is considered in several studies \cite{liu2015qos}. A QoS-aware selection approach is proposed using a multi-dimensional time series \cite{ye2016long}. The proposed approach selects providers based on the consumer's economic models. A qualitative approach is proposed using CP-nets for the long-term selection \cite{mistry2016qualitative}. A QoS-aware approach is proposed to select and compose long-term cloud services based on three meta-heuristic approaches \cite{liu2015qos}. These approaches assume that long-term performance is given for the selection. We focus on a realistic environment where providers disclose limited QoS performance information. 

To the best of our knowledge, existing IaaS selection approaches are not applicable when available QoS performance information is \textit{limited} or \textit{absent}. Free trials are important sources of performance-related information. Cloud consumers often run representative applications in free trials and monitor IaaS performance to select providers. \textit{We aim at leveraging free trials for the long-term selection}. However, the long-term selection based on free trials is challenging due to the \textit{short-term trial restrictions} and the long-term performance variability of the cloud environment \cite{fattah2019long}.  




\section{IaaS Signatures}
\label{sec:signature}

We adopt the concept of signature to represent a provider's long-term performance behavior for a service over a fixed period. The term ``signature'' is typically utilized to indicate the characteristics of an entity, work, or a piece of information that represents their identity or uniqueness. The concept of signature is used for different purposes in several domains such as computing, cryptography, and security. For instance, application performance signatures are used for resource capacity planning and performance anomaly detection \cite{mi2008analysis}. 

An IaaS signature is a relative representation of providers' performance over a fixed period for a particular service. The signature indicates a provider's performance trends and seasonality, i.e., how much a provider's performance may increase or decrease in one time compared to another time. For instance, the signature of an IaaS provider may inform that the provider's performance increase by 10\% on weekend nights than regular weekdays. IaaS signature does not provide the consumer with the actual performance of a provider. A consumer would find it challenging to use the signature without performing the trial using its workloads.

We utilize signatures to measure the confidence of trial experience and to discover a provider's long-term performance. First, the trial confidence is determined by a similarity distance between the trial experience and IaaS signatures. The trial experience may be utilized to discover the long-term service performance when the experience has high confidence. If the trial experience has low confidence, we discard the provider based on a predefined threshold. Next, we utilize the signature to estimate the provider's long-term service performance for the consumer's long-term workloads. The optimal provider is selected based on a time series similarity distance between the consumer's expected service performance and providers' predicted service performance.

\subsection{IaaS Signature Representation}

\noindent \textbf{IaaS Signature}: An IaaS signature is a temporal representation of a provider's relative performance change over time for a service. The signature is defined by a set of QoS parameters that are relevant to the service.

The relevant QoS attributes are the key QoS attributes to measure the performance of the service \cite{li2010cloudcmp}. For example, data read/write throughput, and disk latency are the most important QoS attributes for virtual storage services.



We denote the IaaS signature of a provider as $S=\{S_1, S_2,...S_n\}$, where $n$ is the number of QoS attributes in the signature. Each $S_n$ corresponds to a QoS attribute $Q_n$. $S_n$ denotes a time series for $t$ period where $S_n = \{s_{n1},s_{n2},......s_{nt}\}$. Here, $s_nt$ is the relative performance of the provider at the time $t$ for a particular QoS attribute. We use the following representation IaaS signature:

\scriptsize{
\begin{gather}
 S =
  \begin{bmatrix}
   s_{11} & s_{12} & .. & s_{1t} \\
   s_{21} & s_{22} & .. & s_{2t}  \\
  s_{31} & s_{13} & .. & s_{3t}  \\
   .. & .. & ... \\
   s_{n1} & s_{n2} & .. & s_{nt}  \\
   \end{bmatrix}
   \label{eqn:signature}
\end{gather}
}

\normalsize where each row corresponds to the QoS signature of $Q_i$ and each column represents a timestamp $t$.

\subsection{IaaS Signature Generation}

We aim to represent a provider's long-term service performance changes using its signature. A provider's service performance may vary based on several factors such as the degree of resource overbooking, the number of co-tenants and poor network conditions due to the external factors \cite{wang2018testing}. It is difficult to determine what are the factors behind the performance variability over time from the consumer side. However, the changes in performance often exhibit weekly, monthly, or yearly seasonality \cite{iosup2011performance}. Therefore, it may be possible to capture the seasonal performance changes from the experience of past trial users over different times \cite{wang2018testing}. Note that past trial users may not share their experience publicly to protect their privacy, security, and the conflict of interests with the provider \cite{ba2016new}.


It is reasonable to assume that past users may share their trial experience with a \textit{trusted} non-profit organization (TNPO) for a limited period to help new consumers in the selection \cite{van2012trusted}. Examples of such TNPOs are available in public sectors where privacy-sensitive information about individuals needs to be shared to deliver better services. For instance, health research institutes often collect data about individual patients to improve health services. TNPOs are responsible for data \textit{integration} and \textit{distribution} of collective knowledge without revealing individual's privacy-sensitive information. \textit{We assume that past trial users share their experience with a TNPO.} The TNPO generates IaaS signatures based on the aggregated experience of past trial users. 

IaaS signatures can be generated in different ways based on the purpose of the signatures. The purpose of the signature in this work is two-fold. First, we want to ascertain the confidence of the trial experience using the signature. Second, we want to utilize the signature to predict a provider's future performance behavior using the trial experience. We represent the signature in a way that requires less detailed performance information about the provider and the past trial users. We apply a \textit{normalized averaging} method to generate the signature based on the experience of past trial users.

\begin{figure}
    \centering
    \includegraphics[width=.49\textwidth]{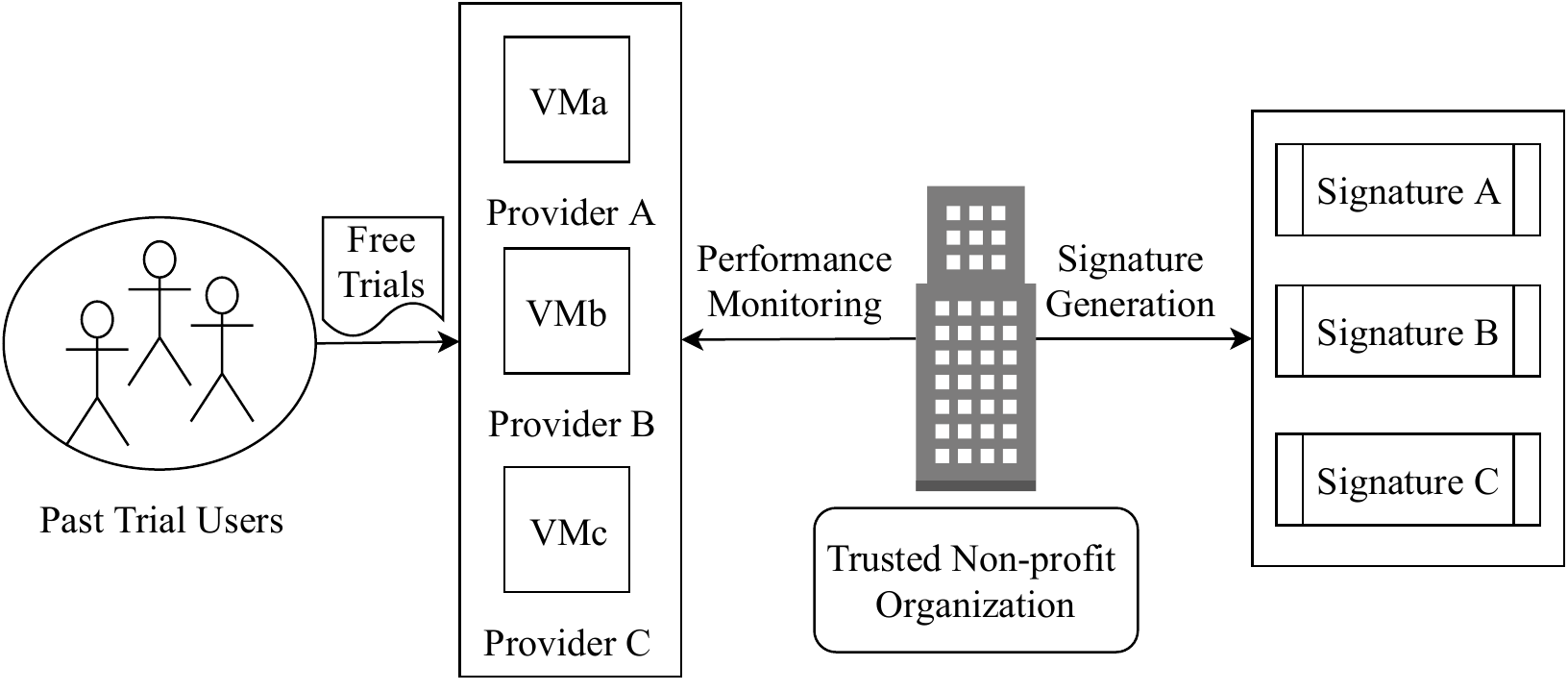}
    \caption{IaaS signature generation}
    \label{fig:tnpo}
    \vspace{-.6cm}
\end{figure}

Let us assume that three IaaS providers ($A$, $B$, and $C$) offer three VMs ($VM_a$, $VM_b$, and $VM_c$) with similar configurations (e.g., capacity, location) for free short-term trials. There exist past users who utilized the trials to find the performance of each VM in different time. Past trial users do not want to share their trial experience publicly. However, each trial user shares their experience with a Trusted Non-profit Organization (TNPO) for a short period (Fig. \ref{fig:tnpo}). The TNPO generates IaaS signatures to identify providers' long-term performance variability for each VM. The TNPO deletes users' experience once the signatures are computed. A signature provides an aggregated view of a provider's long-term performance variability. It is not possible to derive individual trial experience from the signature. As a result, the TNPO does not violate the privacy of past trial users. The IaaS signatures do not contain the provider's actual performance information.

Let us assume that $k$ number of past trial users share their observed trial performance $P_k$ over the period $T$ for a service. We denote $P_k$ as a set of QoS time series where $P_k=\{Q_{1k}, Q_{2k},..., Q_{nk}\}$. Here, $Q_{ik}$ refers to the performance observed by the $k$th consumer for the QoS attribute $Q_i$ over the period $T$. We denote $Q_{ik}$ as $Q_{ik}=\{q_{1ik}, q_{2ik},.., q_{tik}\}$. We perform the following steps to generate the IaaS signature:

\begin{enumerate}[itemsep=0ex, leftmargin=*]
    \item For each QoS attribute $Q_i$, the performance observed by the trial users is collected over time $T$.
    \item At each timestamp $t \in T$, the average performance observed by $k$ number of consumers is measured for each QoS attribute $Q_i$. The average performance over $T$ period is denoted by $\overline{Q_{ik}}$
    \item Each $\overline{Q_{ik}}$ is normalized based on its standard deviation $\sigma (\overline{Q_{ik}})$. The normalized QoS time series groups forms the IaaS signature $S$ over time $T$. 
\end{enumerate}

The value of $s_nt$ at any $t$ represents the relative QoS performance compare to any other time $t'$ in Equation \ref{eqn:signature}. This simple representation of the signature offers two benefits. First, the use of signature becomes easier once a consumer utilizes free trials based on its workloads. The performance for any other time can be found by comparing the ratio between the trial month and other time. Second, signatures can be stored and updated easily over time as it does not require storing detailed information. 

\textit{We assume that the signature provided by the TNPO is accurate and complete for period $T$.} We assume that a provider's signatures does not drastically over these $T$ period. The signature mainly reflects substantial changes in the provider's service performance. The effect of the signature should be visible by most consumers in the trial period unless the provider utilizes an isolated environment. 

\section{Significance-based Trial Scheme}
\label{sec:trial_work}

A consumer needs to evaluate a provider's service performance based on its long-term workloads. \textit{We assume that the consumer's long-term workloads are deterministic, i.e., the expected workloads are known at the time of the selection.} Long-term workloads may be estimated based on the real-world workload traces that can be found in previous activity logs of the consumer. However, it may not possible to run the consumer's entire long-term workloads in a short trial \cite{fattah2019long}. Hence, representative trial workloads need to be generated based on the characteristics of the long-term workloads. 

The first step to generate representative workloads is to determine the workload components (e.g., users, sessions, and applications) and workload parameters. The workload parameters are typically defined by the characteristics of the service requests such as requests arrival times, type of the requests, or resource demands of different types of applications \cite{feitelson2002workload}. We select resource demands per second as the workload parameter without the loss of generality. We denote a consumer's expected workload time series as $w=\{w_1,w_2,...,w_T\}$ over period of time $T$. Here, $w_T$ represents the resource demand at time $T$. Note that it is possible to model other workload parameters as $w$ depending on the service requirements. If multiple workload parameters need to be modeled, we may consider that $w$ has multiple dimensions where each dimension represents a specific workload parameter. The next step is to characterize workloads based on the workload parameters. Workload characterization is usually performed based on statistical analysis such as clustering, specifying dispersion, PCA, and frequency distribution analysis. We use the frequency distribution analysis to characterize the consumer's long-term workloads. Finally, a subset of the long-term workloads is selected as the representative workloads for the trial. The selection criteria are defined based on the characteristics of the long-term workloads. 

A consumer needs to define the selection criteria for the trial workloads carefully. Otherwise, the trial experience may not be helpful for the selection. The trial needs to be performed with the workloads that have the most \textit{significance} to the consumer. We define two types of workload significance based on two workload parameters: a) occurrences, and b) resource consumption. 

\begin{itemize}[itemsep=0ex, leftmargin=*]
    \item \textbf{Frequency-based Significance}: The type of workload that is expected to appear more frequently in the future than any other type of workload is considered significant to the consumer. 
    \item \textbf{Resource Consumption-based Significance}: The type of workload that is expected to demand more resources in the future than any other type of workload is considered significant to the consumer.
\end{itemize}

The workload significance can be defined in terms of other criteria based on various workload parameters. For instance, a consumer may define short-term and long-term requests based on the expected execution time of the requests. We only focus on Frequency-based and Resource consumption-based trial workload generation. Let us assume that the trial period $t$ has $k$ number of timestamps and the consumer's long-term workload has $n$ number of timestamps. We assume that $k<<n$, i.e., the value of $k$ is significantly less than the value of $n$. We need to generate $k$ workloads from $n$ workloads to perform the trial. Algorithm \ref{alg:workload} illustrates the proposed scheme for the trial workload generation. 

\begin{algorithm}


\small

    \caption{Significance-based Trial Scheme}\label{alg:workload}

    \begin{algorithmic}[1]
    
        \State \textbf{Input: }$W$, $t$, $S$
        \State \textbf{Output: }$W_t$

        \State $n \leftarrow size(W)$;
        \State $k \leftarrow length(t)$;
        \State $Uw \leftarrow unique(W)$;
        \State $W_{info} = createArray(size(w))$
        
        \For{each workload $w$ in $Uw$ }
            \State $W_{info}(w).frequency = count(w)$
            \State $W_{info}(w).level = level(w))$
        \EndFor
        \State $W_t = select(W_{info},k,S)$
        \State return $W_t$

    \end{algorithmic}
    
\end{algorithm}

The algorithm \ref{alg:workload} takes the long-term workloads $W$, the trial period $t$, and the significance $S$ as input. The output of the algorithm is $W_t$, which is a subset of $W$. First, the algorithm computes the size of $W$ and the length of $t$. Next, it finds the unique workloads $Uw$ in $W$. An array is then created called $W_{info}$ that stores the \textit{level} of each workload and its \textit{frequency}. The level of a workload defined by the resource consumption of the workloads. For example, if a workload requires 90\% of the CPU units, the level is set to \textit{high} for the workload. The level function is predefined based on the resource capacity. The frequency of each workload is stored based on its number of occurrences in $W$ using the $count$ function. Once the map is created for each workload, a workload \textit{selection} function is used to select $k$ workloads from $n$ workloads using $S$. The value of $S$ determines the significance of the workloads. We use the following three criteria for $S$ to generate the trial workload:

\begin{enumerate}[itemsep=0ex, leftmargin=*]
    \item \textit{Frequency-based Generation (FG):} We select $k$ trial workloads that occur most frequently in $W$. 
    
    \item \textit{Resource Consumption-based Generation (RG):}  We select $k$ trial workloads from $W$ that have maximum resource consumption. 
    \item \textit{Mixed Generation (MG):} We select $k/2$ workloads based on FG method and $k/2$ workloads using RG method.
\end{enumerate}

The selection function can be implemented in different ways based on the workload parameters and the significance. We leave it for the future work to define workload significance using other techniques.

\section{Signature-based IaaS Selection}
\label{sec:selection}

\subsection{Trial Confidence Measure}

The trial confidence is determined using the similarity distance between the IaaS signature and the trial experience. The QoS performance observed in the trial should be normalized before measuring the similarity distance. We measure the similarity distance based on the shape of the signature and the trial experience for each QoS attribute. We decide to use the Pearson Correlation Coefficient (PCC) to measure the trial confidence ($T_{conf}$). The PCC is applied to measure the trial confidence ($T^{Q_i}_{conf}$) for each QoS attributes $Q_i$ as follows:

\scriptsize

\begin{equation}
    T^{Q_i}_{conf} = \frac{\sum_{t=1}^n (q'_t - \Bar{q'}) (q_t - \Bar{q})}{\sqrt{\sum_{t=1}^n (q'_t - \Bar{q'})^2} \sqrt{\sum_{t=1}^n (q_t - \Bar{q})^2}}
    \label{eqn:pearson}
\end{equation}

\begin{figure}
    \centering
    \includegraphics[width=0.35\textwidth]{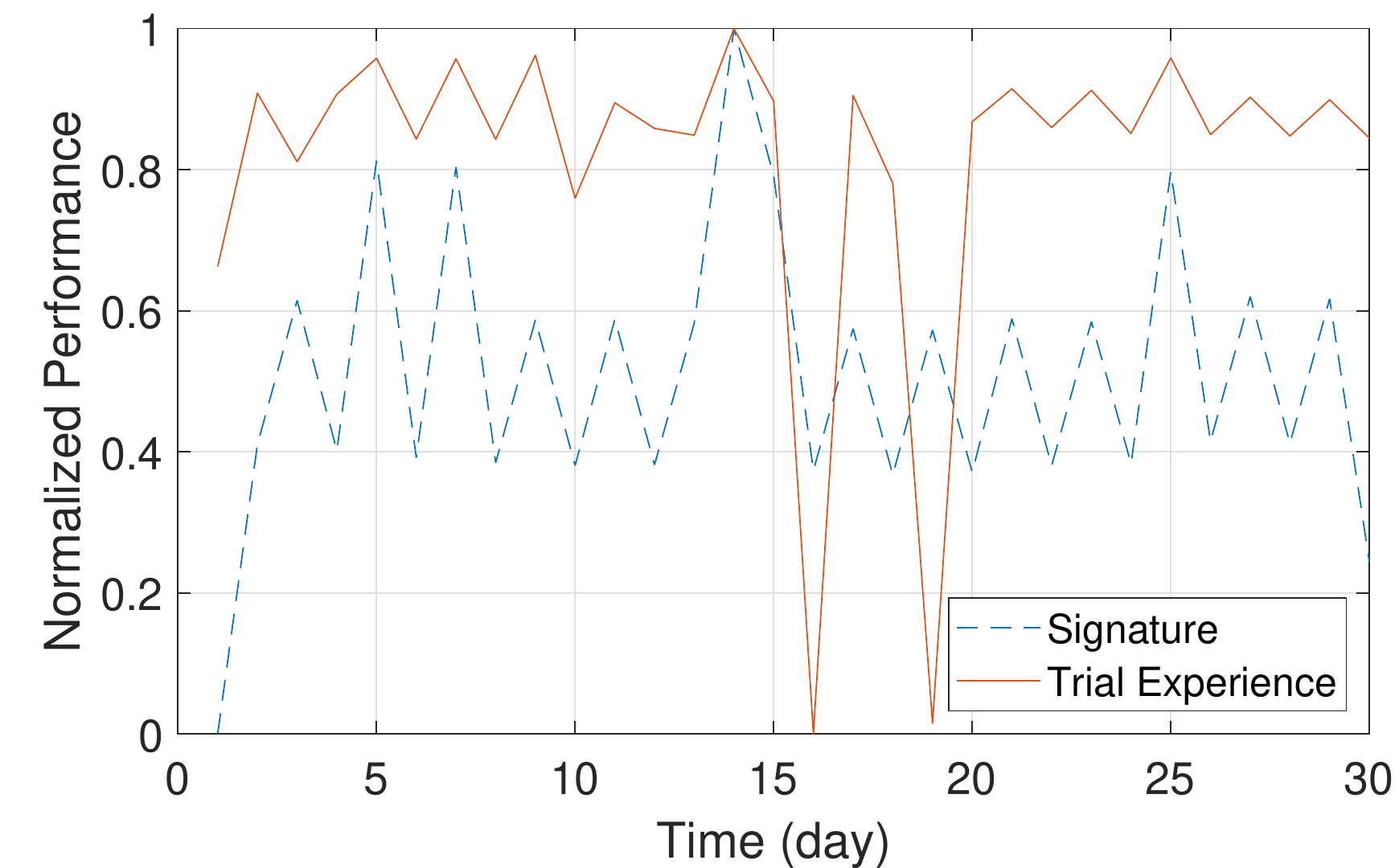}
    \caption{Trial experience matching (CPU throughput)}
    \label{fig:sig_match}
    \vspace{-.6cm}
\end{figure}

\normalsize where $n$ is the trial length, $q'_t$ is the normalized value of the trial performance of $Q_i$ at time $t$. The normalized value of the signature is $q_t$ at time $t$ for the QoS attribute $Q_i$. The mean value of of $Q_i$ is indicated by $\Bar{q'}$. The total confidence calculated by taking the average of all confidence for each QoS attribute by the following equation: 

\scriptsize
\begin{equation}
    T_{conf} = \frac{1}{k}\sum_{i=1}^k T^{Q_i}_{conf}
\end{equation}

\normalsize where $k$ is the total number of QoS attributes. Fig. \ref{fig:sig_match} depicts a plot of normalized trial performance and signature for CPU throughput. The figure shows that the shape of the trial experience is similar to the signature. If the confidence is lower than a predefined threshold (e.g., less than 70\%), the provider is discarded.

\subsection{Signature-based Performance Discovery (SPD)}

We utilize the IaaS signature to measure the provider's service performance beyond the trial period. The first step is to estimate the service performance for the consumer's long-term workloads based on the trial experience. Then, we need to apply the signature to adjust the performance of each type of workload based on the time of its appearance. For example, if a certain type of workload appears in January, then the performance of that type of workload needs to be changed using the signature. We apply algorithm \ref{alg:pred} to discover a provider's service performance beyond the trial period for the consumer's long-term workloads.

\begin{algorithm}
\small
    \caption{Signature-based Performance Discovery}\label{alg:pred}
    \begin{algorithmic}[1]
       
        \State \textbf{Input: }$W$, $W_{trial}$, $P_{trial}$, $S$
        \State \textbf{Output: }$P$ 
        
        \State $TotalTime = length(W)$
        \State $trialLength = length(W_{trial})$
        \State $ S_{trial} = S(1:trialLength) $
        
        \For{each $t$ in $TotalTime$}
            \State $t' =  NearestNeighbor(W(t),W_{trial})$
            \State $TransForm = S(t) / S_{trial}(t') $ 
            \State $P(t) = TransForm* P_{trial}(t') $
            
        \EndFor
        \State return $P$

    \end{algorithmic}
    
\end{algorithm}

Algorithm \ref{alg:pred} takes input the trial workload $W_{trial}$, trial performance $P$, long-term workloads $W$, and IaaS signature $S$. The algorithm returns the long-term performance $P$. First, the algorithm measures the length of $W$ to estimate the total required service time. Then, the trial length is measured based on $W_{trial}$. The part of signature that is applicable for the trial period is taken from $S$ based on the trial length $trialLength$. Next, for each timestamp in the total time, the algorithm needs to measure the performance of the corresponding workload. For each workload at time $t$, the $NearestNeighbor$ function finds the closest workload that can be found in the trial workloads based on resource demand. We use the euclidean distance to measure the similarity between workloads. The function $NearestNeighbor$ returns the timestamp $t'$ that is the timestamp of the closest workload. Next, the transformation factor $TransForm$ is measured by taking the ratio between the signature of the current timestamp $t$ and $t'$. The performance at the current timestamp $P(t)$ is found by multiplying the $P_{trial}(t')$ with the transformation factor $TransForm$. Here, $P_{trial}(t')$ is the performance of the trial workload that is closest to the current workload $W(t)$. 

The performance $P_{trial}$ and $P$ are shown as a one-dimensional time series in the algorithm. However, the algorithm is still applicable if the performance is considered multi-dimensional, i.e., multiple QoS attributes.

\subsection{Long-term IaaS Selection}

The long-term selection is performed based on the consumer's requested performance and the predicted service performance \cite{ye2016long}. First, we normalize the value of QoS attributes based on \textit{Min-Max Feature Scaling} to have the same scale for each QoS attribute using the following equation:

\scriptsize{
\begin{equation}
    Q'_{it} = \frac{Q_{i}(t)-min(Q_i)}{max(Q_i)-min(Q_i)}
    \label{eqn:norm}
\end{equation}
}

\normalsize where $Q_i(t)$ is the value of $Q_i$ at time $t$. $Q_{it}'$ is the normalized value of $Q_{i}(t)$. Equation \ref{eqn:norm} is applied to each QoS time series of the requested and predicted performance. Next, we measure the Root Mean Squared Error (RMSE) distance from the consumer's requested performance and the discovered IaaS performance for each QoS attribute using the following equation:


\scriptsize
\begin{equation}
    RMSE (Q^r_i,Q^p_i) = \sqrt{ \frac{1}{T} \sum_{t=1}^T (Q^r_i(t) - Q^p_i(t))^2 }
\end{equation}




\normalsize where $Q^r_i$ and $Q^p_i$ are the requested and predicted QoS performance of $Q_i$ over time $T$. $Q^r_i(t)$ and $Q^p_i(t)$ denote the requested QoS performance and predicted QoS performance respectively at time $t$. Finally, the rank of each provider is measured by the following equation:

\scriptsize{
\begin{equation}
    Rank (P) = \sum_{i=1}^k RMSE (Q^r_i,Q^p_i)
    \label{eqn:rank}
\end{equation}
}

\normalsize where $Rank(P)$ is the predicted rank of the provider $P$ and $k$ is the total number of QoS attributes. 

\section{Experiments and Results}
\label{sec:exp}

We conduct a set of experiments based on real-world datasets. The proposed SPD approach is compared against the baseline approach, i.e., LPD approach and EQ approach \cite{fattah2019long}. The SPD-based long-term selection is evaluated based on the expected ranking, short-term ranking \cite{ye2016long}, and LPD-based ranking approaches.


\subsection{Experiment Setup}

\begin{table}
\caption{Experiment Variables}
\label{tab:data}
\centering
\begin{tabular}{|l|l|} 
 \hline
 Attribute & Value \\
 \hline
 Total Time & 360 days \\
 Number of Providers & 7 \\
 Trial Period Length & 30 days \\
 Number of Trial Methods & 4 \\
 Trial Month & June \\
 \hline
 \end{tabular}
\vspace{-0.6cm}
\end{table}

\subsubsection{Dataset from Public IaaS Providers}


We run httperf benchmark in Microsoft Azure and Google Compute Engine (GCP) in every 15 minutes for about 1 month. We select Standard A1 v2 and n1-standard-1 types of instances from Azure and GCP respectively. Three instances for each type of VM are installed with similar configurations. Each instance runs a web server that generates a CPU-intensive load (Fibonacci number generator) for each request. The one-month data is divided into 12 partitions. Each partition is considered a one-month data. The signature of each provider is generated using the proposed approach in \ref{sec:signature} based on the data collected from three instances of each provider. 


\subsubsection{Dataset from Private IaaS Providers}

\begin{figure*}[!t]
\vspace{-.7cm}
    \centering
    \subfloat[]{\includegraphics[width=.33\textwidth]{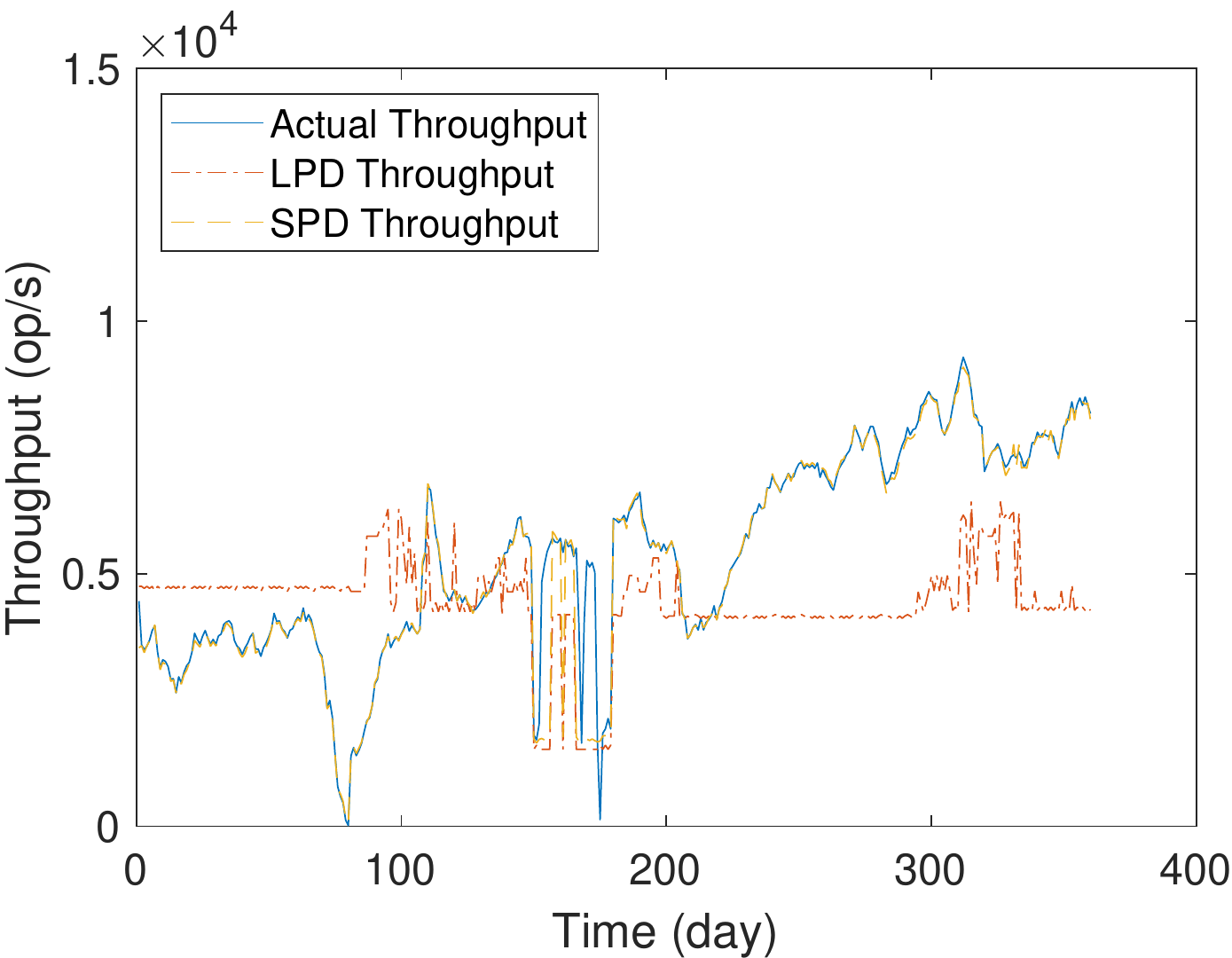}}
    \hfil
    \subfloat[]{\includegraphics[width=.33\textwidth]{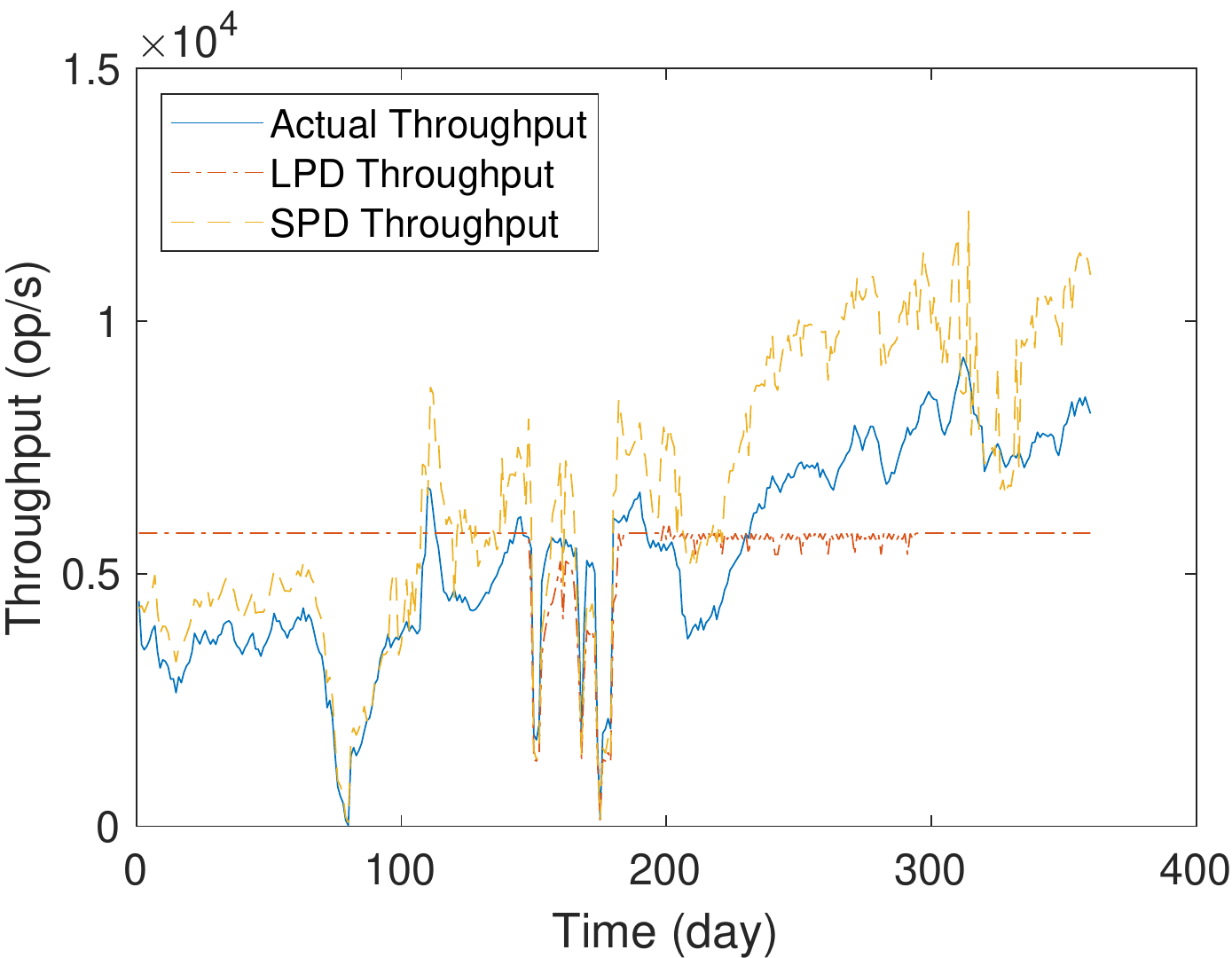}}
    \hfil
    \subfloat[]{\includegraphics[width=.33\textwidth]{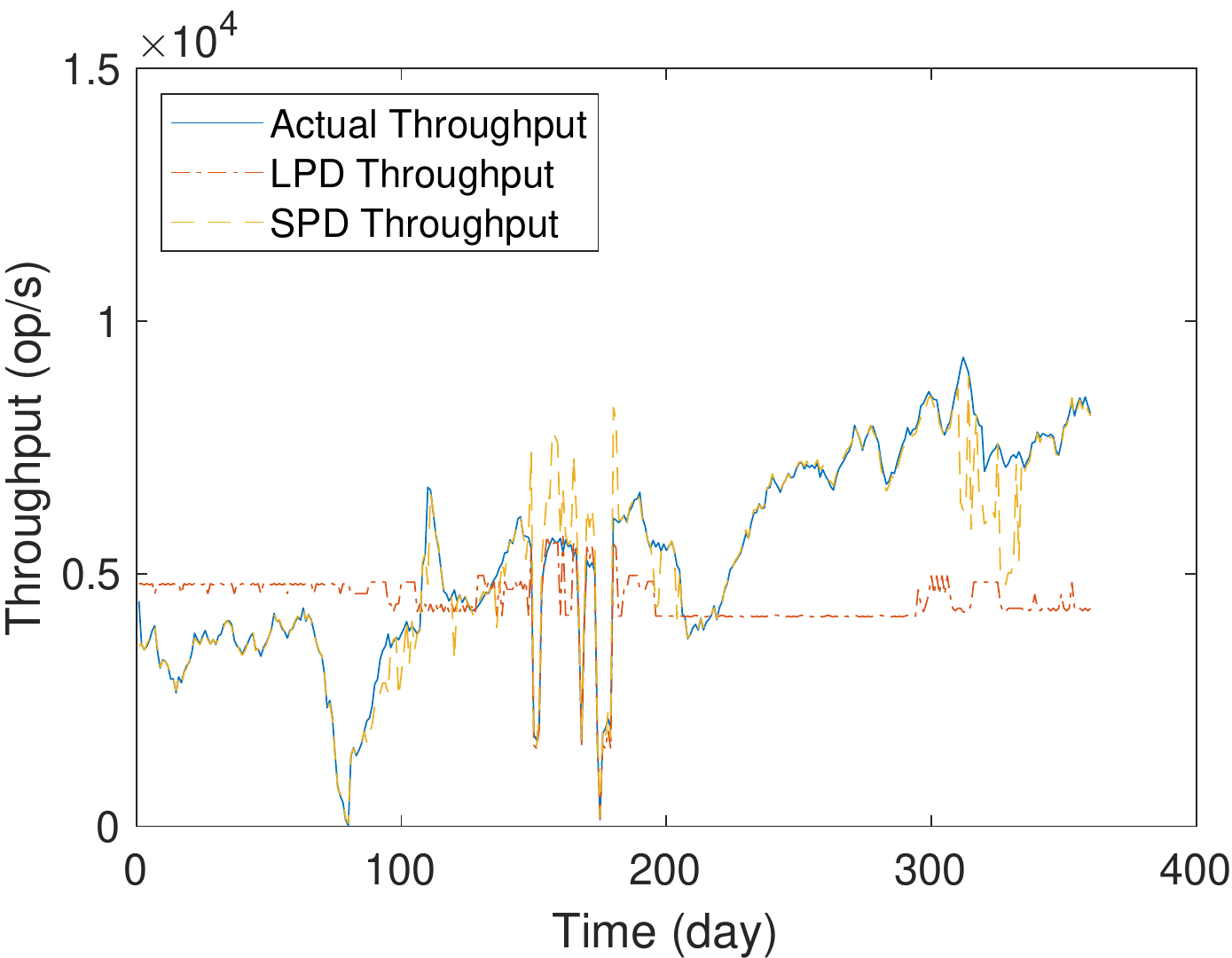}}
    \vspace{-.4cm}
    \hfil
    \subfloat[]{\includegraphics[width=.33\textwidth]{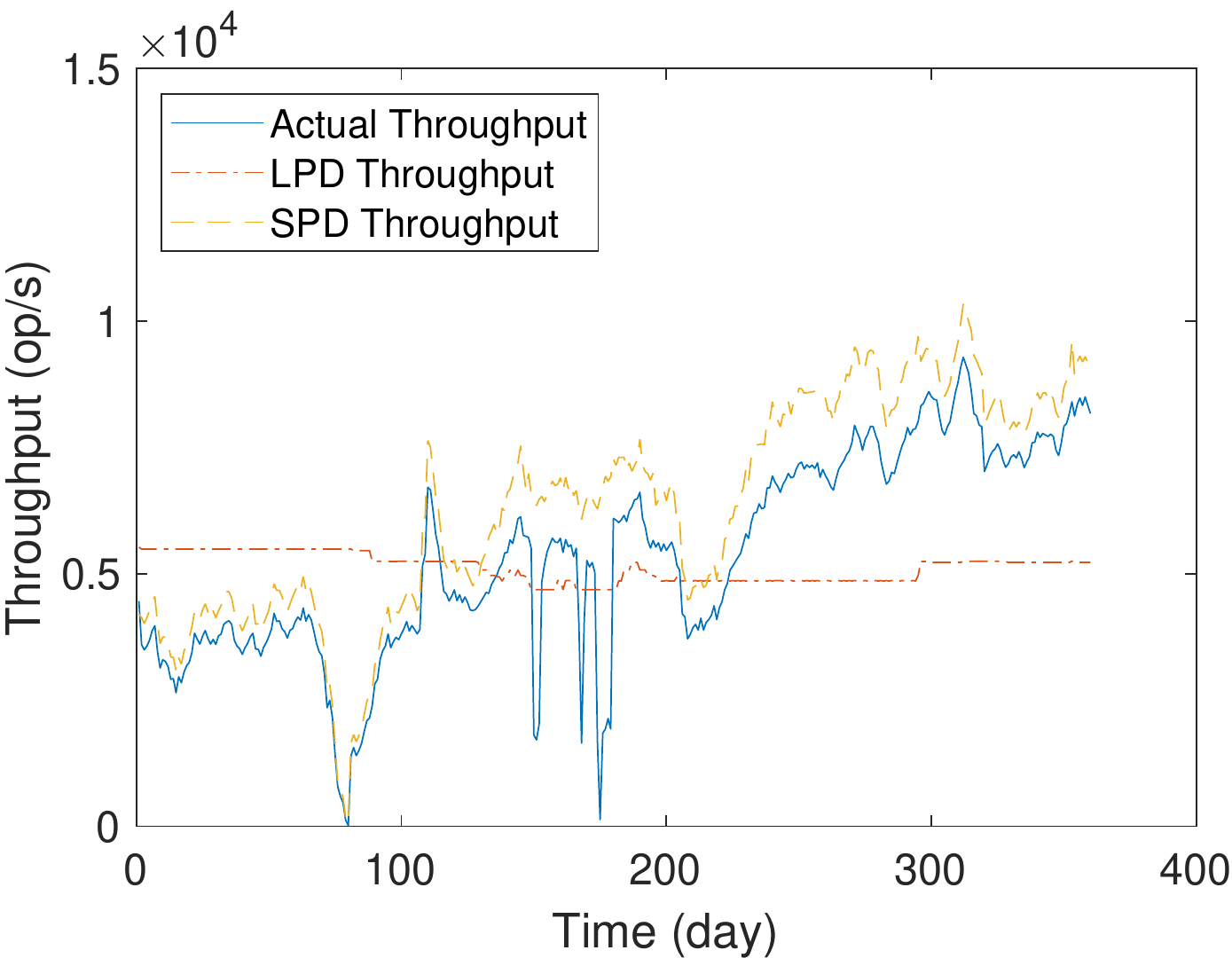}}
    \hfil
    \subfloat[]{\includegraphics[width=.33\textwidth]{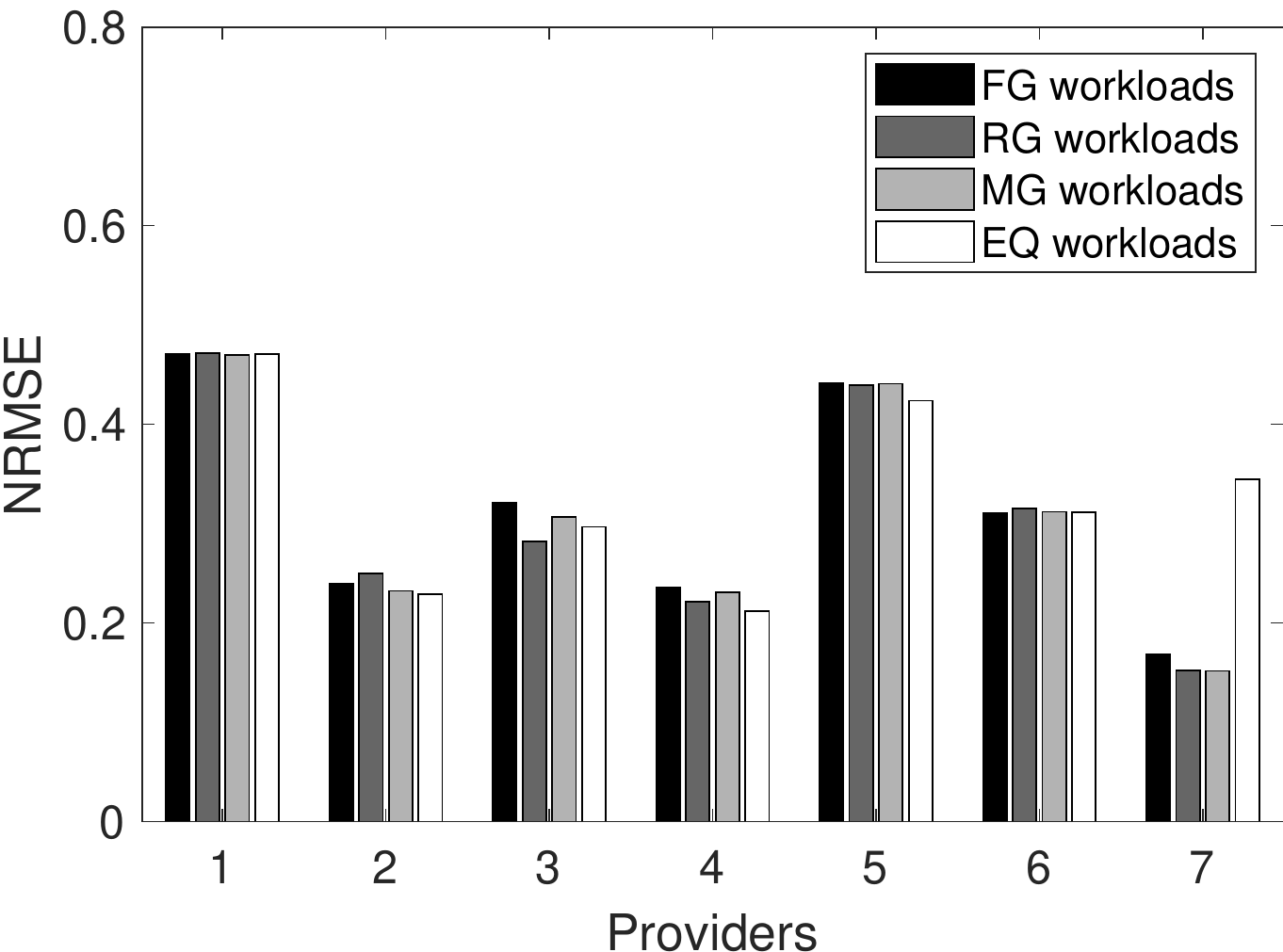}}
    \hfil
    \subfloat[]{\includegraphics[width=.33\textwidth]{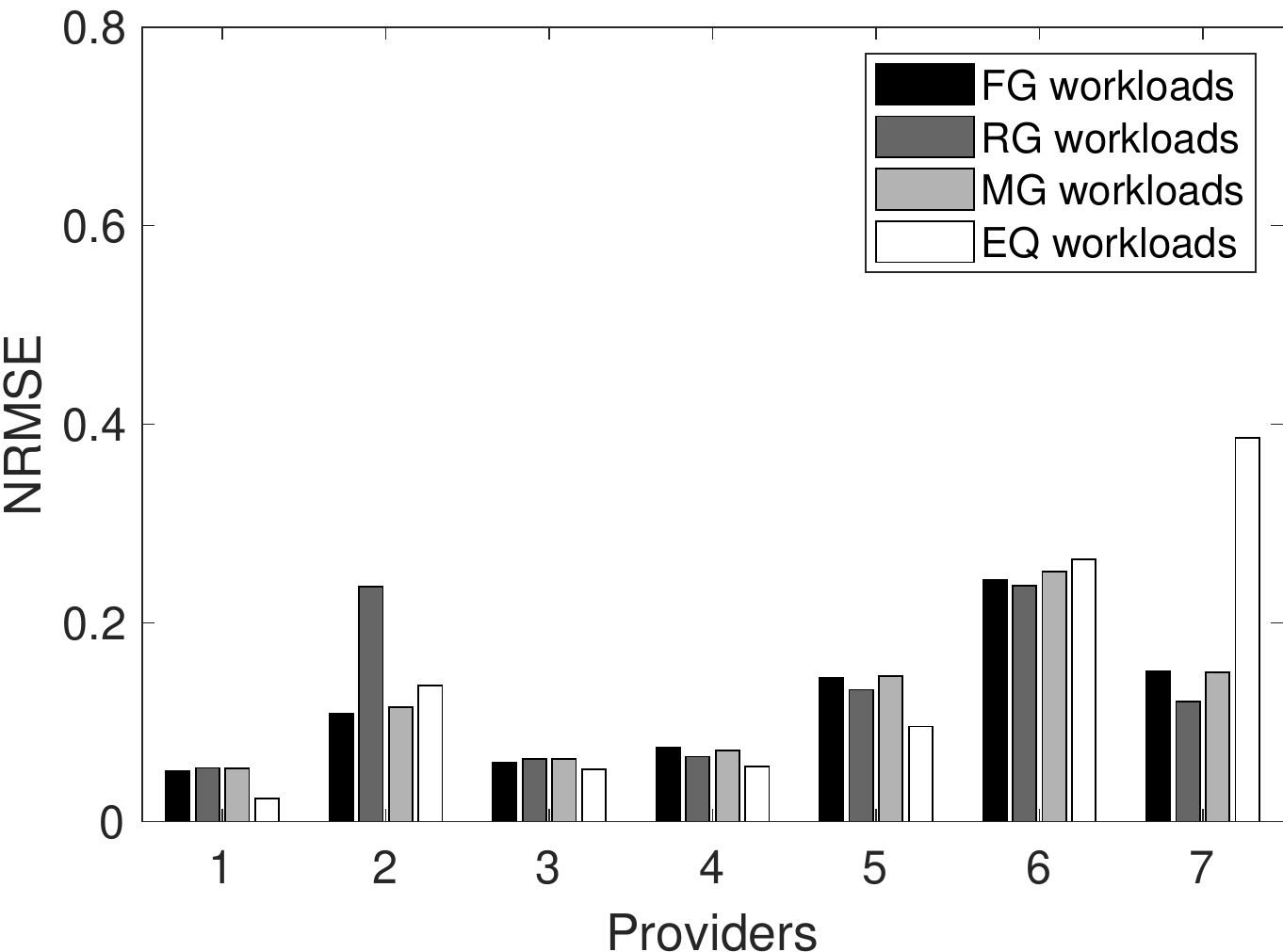}}

    \caption{Long-term throughput prediction (a) FG workloads (b) RG workloads (c) MG workloads (d) EQ workloads (e) LPD approach error (f) SPD approach error}
    \label{fig:qos_sig}
    \vspace{-.6cm}
\end{figure*}

We utilize publicly available  Eucalyptus IaaS workload\footnote{\url{https://sites.cs.ucsb.edu/~rich/workload/}} to generate long-term consumer workloads. It contains about 34 days of workload data. We generate 360 days of workload data for each consumer based average workload per day. The long-term performance of 5 private IaaS providers is generated from benchmark results published SPEC Cloud IaaS 2016\footnote{\url{https://www.spec.org/}}. First, we map each unique workload of the cluster to unique performance value of the benchmark results. We consider the map as a baseline performance for the workload. Next, we build long-term performance profiles for the providers where each provider shows different performance behavior based on the workloads and time. We run the workloads of each consumer on five providers for the long-term period and a short-term trial period to discover the corresponding performances of each provider. The experiment variables are shown in Table \ref{tab:data}.

\subsubsection{Baseline Approach} We define a Long-term Performance Discovery (LPD) approach as the baseline approach to evaluate the proposed SPD approach. The trial experience contains a subset of long-term workloads and corresponding performance. We generate the performance of the long-term workloads for each provider based on the consumer trial experience. For each workload $w_l$ in the long-term workloads, we find a workload $w_t$ in the trial workload, where $w_l$ and $w_t$ have similar resource consumption. We consider the performance of $w_l$ is equivalent to $w_t$.  

\subsubsection{Equivalence Partitioning-based Approach} An equivalence partitioning-based (EQ) approach is proposed in \cite{fattah2019long} where the consumer's long-term workload is partitioned based on the number of available VMs in the free trial period. Then, workloads of each partition are compressed within one day assuming that the performance of the provider does not change considerably within a day. Each VM runs the same workload for the trial period to understand the performance variability.

\subsection{Evaluation of Long-term Performance Discovery}

Fig. \ref{fig:qos_sig} shows the results of the long-term IaaS performance discovery. Fig. \ref{fig:qos_sig}(a), (b), (c), and (d) show the predicted CPU throughput of a provider using the FG, RG, MG, and EQ trial schemes. Each figure shows the LPD throughput, SPD throughput, and actual throughput. The predicted performance using the LPD approach exhibits similar behavior in each figure. The LPD predicted performance cannot capture the temporal performance shifts. It is noticeable that the predicted performance remains on the same performance level of the trial month (151-180 days). The LPD approach can be useful to predict the performance of the providers that provide services with good performance isolation. The SPD approach predicts the throughput more accurately compared to the LPD approach as shown in each figure (Fig. \ref{fig:qos_sig}(a), (b), (c), and (d) ). The SPD approach utilizes the shape of the signature to estimate long-term IaaS performance. Hence, the predicted performance has a similar shape to the signature. Fig. \ref{fig:qos_sig}(e) and (f) shows the accuracy of the predicted performance using normalize RMSE (NRMSE) distances for 7 providers. Provider 6 and 7 are the public IaaS provider and the rest are private IaaS providers. The accuracy of the SPD approach (Fig. \ref{fig:qos_sig}(f)) is considerably higher than the LPD approach (Fig. \ref{fig:qos_sig}(e)). 


\subsection{Effect of Trial Schemes in Performance Discovery} 

The effects of different trial schemes are noticeable in Fig. \ref{fig:qos_sig}(e) and (f). The LPD approach exhibits less performance variability for different trial schemes as it does not consider the provider's long-term performance variability. The prediction accuracy of the SPD approach varies considerably based on the selected trial scheme. The RG scheme-based SPD approach (Fig. \ref{fig:qos_sig}(f)) shows the lowest accuracy compared to the other approaches (Fig. \ref{fig:qos_sig}(f)). It is due to the characteristics of the consumer's long-term workloads. The RG scheme mainly selects resource-intensive (i.e., requires high resource usage) workloads similar to the traditional \textit{load} and \textit{stress} testing based approaches. Hence, traditional load and stress testing techniques may not provide good accuracy for long-term performance discovery.

The FG scheme-based SPD approach shows the maximum estimation accuracy compared to the other trial schemes (Fig. \ref{fig:qos_sig} (f)). The reason is that it utilizes most frequently occurred workloads in the consumer's long-term workloads. The maximum number of workloads are tested in this scheme. The estimation errors for the MG and EQ scheme remain in between the FG and RG schemes. MG scheme is built using the FG and RG scheme. As a result, the NRMSE for the MG scheme is in between FG and RG schemes. The EQ scheme shows poor performance for public providers. The reason is that EQ scheme depends on the number of available VMs to run the experiments.

\subsection{Evaluation of IaaS Ranking}

\begin{table}
\centering
\caption{Ranking of IaaS Providers}
\begin{tabular}{|l|c|l|}
\hline
  Rankings & Orders  \\
 \hline
Expected  & p1 $<$ p4 $<$ p2 $<$ p3 $<$ p5 $<$ p6 $<$ p7 \\
 \hline
Short-term  & p1 $<$ p6 $<$ p7 $<$ p3 $<$ p4 $<$ p2 $<$ p5 \\
 \hline
LPD  & p2 $<$ p4 $<$ p5 $<$ p3 $<$ p1 $<$ p6 $<$ p7 \\
 \hline
SPD  & p1 $<$ p4 $<$ p3 $<$ p2 $<$ p5 $<$ p6 $<$ p7 \\
 \hline
 \end{tabular}
\label{tab:ranks}
\vspace{-.5cm}
\end{table}

The ranking of the provider based on different approaches is shown in Table \ref{tab:ranks}. We measure the \textit{expected ranking} of the providers to evaluate the proposed selection approach. The expected ranking is computed based on the NRMSE distance between the consumer's throughput requirement and a provider's actual throughput. We rank the providers based on three approaches using the FG scheme. First, we rank the providers based on the short-term trial experience. The short-term ranking cannot rank the providers correctly compared to the expected ranking. Therefore, the short-term selection approach is not applicable for the long-term period. Next, we rank the providers based on the predicted performance using the LPD approach, which does not rank most providers correctly. Hence, the selection based on the trial experience without considering the long-term performance may lead to wrong provider selection. Finally, we rank the providers based on the predicted performance using the SPD approach that ranks most providers correctly. 




\section{Conclusion}
\label{sec:con}

We introduce a novel approach to select the optimal IaaS service according to a consumer's long-term QoS requirements. The proposed approach leverages free trials and IaaS signatures to discover long-term service performance of IaaS providers. The experiment results using the real-world datasets show that the proposed SPD approach effectively discovers long-term service performance using different trial schemes. We conclude that the selection of an appropriate trial scheme plays an important role in the long-term performance discovery. The results also confirm that the proposed approach ranks the IaaS services effectively using the IaaS signatures and the consumer's trial experience. We focus on the deterministic workloads in this work. In the future, we will explore the long-term IaaS selection for the stochastic workloads.


\section{Acknowledgement}

This research was partly made possible by DP160103595 and LE180100158 grants from the Australian Research Council. The statements made herein are solely the responsibility of the authors.

\bibliographystyle{IEEEtran}
\bibliography{IEEEabrv,Main}

\begin{thebibliography}{10}
\providecommand{\url}[1]{#1}
\csname url@samestyle\endcsname
\providecommand{\newblock}{\relax}
\providecommand{\bibinfo}[2]{#2}
\providecommand{\BIBentrySTDinterwordspacing}{\spaceskip=0pt\relax}
\providecommand{\BIBentryALTinterwordstretchfactor}{4}
\providecommand{\BIBentryALTinterwordspacing}{\spaceskip=\fontdimen2\font plus
\BIBentryALTinterwordstretchfactor\fontdimen3\font minus
  \fontdimen4\font\relax}
\providecommand{\BIBforeignlanguage}[2]{{%
\expandafter\ifx\csname l@#1\endcsname\relax
\typeout{** WARNING: IEEEtran.bst: No hyphenation pattern has been}%
\typeout{** loaded for the language `#1'. Using the pattern for}%
\typeout{** the default language instead.}%
\else
\language=\csname l@#1\endcsname
\fi
#2}}
\providecommand{\BIBdecl}{\relax}
\BIBdecl

\bibitem{iosup2014iaas}
A.~Iosup, R.~Prodan, and D.~Epema, ``Iaas cloud benchmarking: approaches,
  challenges, and experience,'' in \emph{Cloud Computing for Data-Intensive
  Applications}.\hskip 1em plus 0.5em minus 0.4em\relax Springer, 2014, pp.
  83--104.

\bibitem{ye2016long}
Z.~Ye, S.~Mistry, A.~Bouguettaya, and H.~Dong, ``Long-term qos-aware cloud
  service composition using multivariate time series analysis,'' \emph{IEEE
  TSC}, vol.~9, no.~3, pp. 382--393, 2016.

\bibitem{mistry2018metaheuristic}
S.~Mistry, A.~Bouguettaya, H.~Dong, and A.~K. Qin, ``Metaheuristic optimization
  for long-term iaas service composition,'' \emph{IEEE TSC}, vol.~11, no.~1,
  pp. 131--143, 2018.

\bibitem{fattah2019long}
S.~M.~M. Fattah, A.~Bouguettaya, and S.~Mistry, ``Long-term iaas provider
  selection using short-term trial experience,'' in \emph{ICWS}.\hskip 1em plus
  0.5em minus 0.4em\relax IEEE, 2019, pp. 304--311.

\bibitem{yu2010computing}
Q.~Yu and A.~Bouguettaya, ``Computing service skylines over sets of services,''
  in \emph{IEEE ICWS}.\hskip 1em plus 0.5em minus 0.4em\relax IEEE, 2010, pp.
  481--488.

\bibitem{fattah2018cp}
S.~M.~M. Fattah, A.~Bouguettaya, and S.~Mistry, ``A cp-net based qualitative
  composition approach for an iaas provider,'' in \emph{WISE}.\hskip 1em plus
  0.5em minus 0.4em\relax Springer, 2018, pp. 151--166.

\bibitem{wang2018testing}
W.~Wang, N.~Tian, S.~Huang, S.~He, A.~Srivastava, M.~L. Soffa, and L.~Pollock,
  ``Testing cloud applications under cloud-uncertainty performance effects,''
  in \emph{ICST}.\hskip 1em plus 0.5em minus 0.4em\relax IEEE, 2018, pp.
  81--92.

\bibitem{iosup2011performance}
A.~Iosup, N.~Yigitbasi, and D.~Epema, ``On the performance variability of
  production cloud services,'' in \emph{CCGrid}.\hskip 1em plus 0.5em minus
  0.4em\relax IEEE, 2011, pp. 104--113.

\bibitem{feitelson2002workload}
D.~G. Feitelson, ``Workload modeling for performance evaluation,'' in
  \emph{IFIP}.\hskip 1em plus 0.5em minus 0.4em\relax Springer, 2002, pp.
  114--141.

\bibitem{li2010cloudcmp}
A.~Li, X.~Yang, S.~Kandula, and M.~Zhang, ``Cloudcmp: comparing public cloud
  providers,'' in \emph{IMC}.\hskip 1em plus 0.5em minus 0.4em\relax ACM, 2010,
  pp. 1--14.

\bibitem{liu2015qos}
S.~Liu, Y.~Wei, K.~Tang, A.~K. Qin, and X.~Yao, ``Qos-aware long-term based
  service composition in cloud computing,'' in \emph{CEC}.\hskip 1em plus 0.5em
  minus 0.4em\relax IEEE, 2015, pp. 3362--3369.

\bibitem{jayasinghe2012expertus}
D.~Jayasinghe, G.~Swint, S.~Malkowski, J.~Li, Q.~Wang, J.~Park, and C.~Pu,
  ``Expertus: A generator approach to automate performance testing in iaas
  clouds,'' in \emph{CLOUD}.\hskip 1em plus 0.5em minus 0.4em\relax IEEE, 2012,
  pp. 115--122.

\bibitem{mistry2016qualitative}
S.~Mistry, A.~Bouguettaya, H.~Dong, and A.~Erradi, ``Qualitative economic model
  for long-term iaas composition,'' in \emph{ICSOC}.\hskip 1em plus 0.5em minus
  0.4em\relax Springer, 2016, pp. 317--332.

\bibitem{mi2008analysis}
N.~Mi, L.~Cherkasova, K.~Ozonat, J.~Symons, and E.~Smirni, ``Analysis of
  application performance and its change via representative application
  signatures,'' in \emph{NOMS}.\hskip 1em plus 0.5em minus 0.4em\relax IEEE,
  2008, pp. 216--223.

\bibitem{ba2016new}
M.~N. Ba-Hutair and I.~Kamel, ``A new scheme for protecting the privacy and
  integrity of spatial data on the cloud,'' in \emph{BigMM}.\hskip 1em plus
  0.5em minus 0.4em\relax IEEE, 2016, pp. 394--397.

\bibitem{van2012trusted}
S.~W. van~den Braak, S.~Choenni, R.~Meijer, and A.~Zuiderwijk, ``Trusted third
  parties for secure and privacy-preserving data integration and sharing in the
  public sector,'' in \emph{DGO}.\hskip 1em plus 0.5em minus 0.4em\relax ACM,
  2012, pp. 135--144.

\end{thebibliography}

\end{document}